\documentclass[useAMS,usenatbib]{mn2e} 
\usepackage{graphicx}
\usepackage{amsmath}
\usepackage{amssymb} 
\title{Origin of the DAMPE 1.4 TeV peak} 
\author[Chan]{Man Ho Chan \thanks{chanmh@eduhk.hk}, Chak Man Lee
\\ Department of Science and Environmental Studies, The Education University of Hong Kong, Tai Po, Hong Kong}

\begin{document}

\date{Accepted XXXX, Received XXXX}

\pagerange{\pageref{firstpage}--\pageref{lastpage}} \pubyear{XXXX}

\maketitle

\label{firstpage}

\date{\today}

\begin{abstract}
Recent accurate measurements of cosmic ray electron flux by the Dark Matter Particle Explorer (DAMPE) reveal a sharp peak structure near 1.4 TeV, which is difficult to explain by standard astrophysical processes. In this letter, we propose a simple model that the enhanced dark matter annihilation via the $e^+e^-$ channel and with the thermal relic annihilation cross section around the current nearest black hole (A0620-00) can satisfactorily account for the sharp peak structure. The predicted dark matter mass is $\sim 1.5-3$ TeV.
\end{abstract}


\section{Introduction}
Observations of galaxies and galaxy clusters reveal the existence of dark matter. Many theoretical models suggest that dark matter particles can self-annihilate to give electrons, positrons, neutrinos, etc. If dark matter particles are thermal relic particles, the predicted thermal relic annihilation cross section is $\sigma v=2.2 \times 10^{-26}$ cm$^3$ s$^{-1}$ for dark matter mass $m \ge 10$ GeV \citep{Steigman}. 

In the past decade, observations of HEAT \citep{Beatty}, PAMELA \citep{Adriani} and AMS-02 \citep{Aguilar,Aguilar2} reported some excess positron emissions in our Milky Way. These excess positrons can be explained by dark matter annihilation \citep{Boudaud,Mauro}. However, recent analyses of gamma-ray and radio observations put some stringent constraints for annihilating dark matter \citep{Ackermann,Albert,Chan2,Egorov,Chan3,Chan4,Chan5}, which give strong tensions for the dark matter interpretation of the positron excess. Besides, many studies start to point out that pulsars' emission can satisfactorily account for the positron excess in our galaxy \citep{Linden,Delahaye2}.

Recently, accurate measurements of electron/positron flux by the Dark Matter Particle Explorer (DAMPE) have been published \citep{Ambrosi}. Surprisingly, a sharp peak structure near 1.4 TeV is discovered. This sharp peak likely originates from nearby source because the energy of electrons and positrons would cool down severely after their propagation \citep{Pan}. Various studies have been performed to explain this sharp peak structure, including dark matter annihilation and dark matter decay from nearby dark matter subhalo or ultra-compact micro halos \citep{Pan,Jin,Athron,Huang,Gao,Niu,Yang,Ding,Ge}. However, the required annihilation cross section obtained is much larger than the thermal relic annihilation cross section and the decay rate obtained is difficult to verify independently. Moreover, the details of the nearby dark matter subhalo are quite uncertain. We still do not have any strong evidence showing the existence of dark matter subhalos in our galaxy.

In this letter, we propose a new simple model which can satisfactorily explain the sharp peak structure. Standard gravitational theory suggests that dark matter distribution would be influenced by the black holes nearby \citep{Gondolo}. The dark matter density profile would form a mini-spike which can enhance dark matter annihilation rate \citep{Zhao,Bertone, Bertone2,Lacroix,Chan}. We find that the electrons and positrons emitted from the current nearest black hole (A0620-00) due to the enhanced dark matter annihilation can account for the sharp peak in the energy spectrum. 

\section{The density spike model}
The nearest black hole from us (A0620-00) has been discovered for a long time \citep{Cantrell}. Based on the latest data, the mass of the black hole is $M_{\rm BH}=6.61 \pm 0.25M_{\odot}$ and the distance is $d=1.06\pm 0.12$ kpc \citep{Cantrell}. If the black hole grew adiabatically, the dark matter density would be enhanced in a region corresponding to the sphere of influence of the black hole \citep{Lacroix2}. A dark matter spike would be produced and the dark matter density would go like $r^{-\gamma_{\rm sp}}$ with $\gamma_{\rm sp}=1.5-2.25$ \citep{Gnedin}. The dark matter density is given by \citep{Lacroix}
\begin{equation}
\rho(r)= \left \{ \begin{array}{lll}
0       & {\ \ r \le 2R_S } \\ &\\
\rho_{\rm sat}       & {\ \ 2R_S < r \le R_{\rm sat} } \\ &\\
\rho_0 \left(\frac{r}{R_{\rm sp}} \right)^{-\gamma_{\rm sp}},       & {\ \ R_{\rm sat} < r \le R_{\rm sp} } \end{array} \right.
\end{equation} 
where $\rho_{\rm sat}=m/(\sigma vt_{\rm BH})$ is the saturation density with $t_{\rm BH}$ is the age of the black hole, and $R_{\rm sat}=R_{\rm sp}(\rho_{\rm sat}/\rho_0)^{-1/\gamma_{\rm sp}}$. The normalization density $\rho_0$ can be usually determined by assuming the mass inside the spike $M_{\rm sp} \approx M_{\rm BH}$ \citep{Lacroix}. This gives $\rho_0 \approx (3-\gamma_{\rm sp})M_{\rm BH}/(4\pi R_{\rm sp}^3)$. In the following, we assume that it is an adiabatic mini-spike, i.e. $\gamma_{\rm sp}=9/4$ \citep{Gnedin}.

The total integrated electron emission rate (in GeV$^{-1}$ s$^{-1}$) for the mini-spike is
\begin{equation}
\Phi_{\rm sp}= \frac{\sigma v}{2m^2} \int_0^{R_{\rm sp}} 4\pi r^2 \rho^2(r)dr \frac{dN_i}{dE},
\end{equation}
where $dN_i/dE$ is the electron spectrum per dark matter annihilation, which can be obtained in \citet{Cirelli}. By combining Eqs.~(1) and (2), the electron plus positron flux detected is given by \citep{Lacroix}
\begin{equation}
\begin{aligned}
\phi_{\rm sp}
=& 9.0 \times 10^{-8} m_{\rm TeV}^{-4/3} \left( \frac{d}{1.06~\rm kpc} \right)^{-2} \left( \frac{M_{\rm BH}}{10M_{\odot}} \right)^{4/3}  
\\
& \times \left( \frac{R_{\rm sp}}{10^{-3}~\rm pc} \right)^{-1} \left( \frac{\sigma v}{2.2 \times 10^{-26}~\rm cm^3~s^{-1}} \right)^{1/3} 
\\
& \times \left(\frac{t_{\rm BH}}{10^{10}~{\rm yr}} \right) \frac{dN}{dE}~{\rm GeV^{-1}~cm^{-2}~s^{-1}}, 
\end{aligned}
\end{equation}
where $m_{\rm TeV}$ is the dark matter mass in TeV and $dN/dE$ is the energy spectrum of electrons or positrons after propagation. 

By taking $M_{\rm BH}=6.61M_{\odot}$ and extrapolating the relation of black hole mass and velocity dispersion ($M_{\rm BH}-\sigma_*$ relation) \citep{Tremaine}, we can get $R_{\rm sp} \approx GM_{\rm BH}/\sigma_*^2=3.08 \times 10^{-3}$ pc. Since the black hole is located in the thin disk of our galaxy, the age of the black hole $t_{\rm BH}$ should be smaller than the age of the thin disk $8.8 \pm 1.7$ Gyr \citep{Peloso}. In the following, we assume a reasonable value $t_{\rm BH} \approx 5$ Gyr. Using the thermal relic annihilation cross section $\sigma v=2.2 \times 10^{-26}$ cm$^3$ s$^{-1}$ and the distance of the black hole $d=1.06$ kpc, we get
\begin{equation}
\phi_{\rm sp}=8.4 \times 10^{-9}m_{\rm TeV}^{-4/3} \frac{dN}{dE}~{\rm GeV^{-1}~cm^{-2}~s^{-1}}.
\end{equation}

During propagation, the electrons and positrons produced from dark matter annihilation would mainly cool down by synchrotron radiation and inverse Compton scattering. The cooling rate is $b(E)=b_0E^{-2}$, where $b_0$ is a constant which depends on the energy density of the environment and $E$ is the energy of the electrons or positrons in GeV. Taking the magnetic field strength $B \approx 2.1~\mu$G at solar location \citep{Cirelli} (magnetic energy density $\omega_{\rm B}=0.106$ eV/cm$^3$), the cosmic microwave background radiation energy density $\omega_{\rm cmb}=0.25$ eV/cm$^3$ and the optical-infrared radiation energy density $\omega_{\rm opt}=0.5$ eV/cm$^3$ \citep{Atoyan}, the total energy density is $\omega=0.856$ eV/cm$^3$, which corresponds to $b_0=8.7\times 10^{-17}$ GeV s$^{-1}$. 

The diffusion length of an electron with initial energy $E_i$ and final energy $E_f$ is given by \citep{Fornengo}
\begin{equation}
d_f= \left(4 \int_{E_f}^{E_i} \frac{K_0E^{\delta}}{b(E)}dE \right)^{1/2},
\end{equation}
where $K_0$ is the diffusion coefficient and $\delta$ is the diffusion index. Recent studies obtain $K_0$ and $\delta$ for three different benchmark propagation models (MIN, MED and MAX) (see Table 1) \citep{Fornengo,Delahaye}. Putting $d_f=d=1.06$ kpc and using the energy spectrum of dark matter annihilation $dN_i/dE$ for the initial energy $E_i$, we can obtain the energy spectrum for the final energy $E_f$ after propagation (see Fig.~1).

\section{Fitting with the DAMPE data}
The spectra of cosmic ray electrons and positrons were measured by DAMPE with great accuracy \citep{Ambrosi}. The DAMPE data are dominated by the background emission, which includes three different sources: primary electrons generated by astrophysical sources, secondary electrons/positrons and pulsars. The empirical fitted spectrum (in GeV$^{-1}$ m$^{-2}$ s$^{-1}$ sr$^{-1}$) for primary electrons/positrons, secondary electron/positrons and pulsars are respectively modeled by \citep{Pan}
\begin{equation}
\phi_{e^-}=21.45E^{-0.86}[1+(E/3.38)^{2.38}]^{-1}\exp[-(E/1296.60)^4],
\end{equation} 
\begin{equation}
\phi_{e^{\pm}}=0.88E^{-0.47}[1+(E/2.70)^{2.80}]^{-1},
\end{equation}
and
\begin{equation}
\phi_s=1.35E^{-2.47}\exp(-E/1801.90).
\end{equation}
The total flux spectrum of the background cosmic ray electrons and positrons can be described by \citep{Pan}
\begin{equation}
\phi_{\rm back}=\phi_{e^-}+1.6\phi_{e^{\pm}}+2\phi_s.
\end{equation}
Here, the factor 1.6 corresponds to the extra positrons produced due to proton-proton collision \citep{Kamae}. 

Therefore, the total flux spectrum is $\phi=\phi_{\rm back}+\phi_{\rm sp}$. Dark matter annihilation via the $e^+e^-$ channel can give a very narrow electron/positron energy spectrum which peaks at $E=m$. For $d=1.06$ kpc and the spectral peak $E_f=1.4$ TeV, we get $E_i=m=2.75$ TeV with the MED model parameters (see Fig.~1). In other words, a large amount of electrons and positrons produced from dark matter annihilation around the black hole with initial energy 2.75 TeV cool down to 1.4 TeV after travelling by a distance of 1.06 kpc.

In Fig.~2, we fit the DAMPE data with our model (MED model parameters with $m=2.75$ TeV) and we find that it gives an excellent agreement with the data. We also test for the MIN and MAX model parameters (see Fig.~3). The dark matter mass predicted for these two models are 15 TeV and 1.7 TeV respectively. However, only the MAX model can produce a good agreement with the data as the flux $\phi_{\rm sp}$ for the MIN model is too small. In Fig.~2, we also compare our MED model fit with the Fermi-LAT electron/positron spectrum obtained in \citet{Abdollahi}. Our result generally agrees with the Fermi-LAT spectrum, only except for the data points at 1.7 TeV and 1.9 TeV. Nevertheless, if we consider the systematic uncertainty on the energy reconstruction (see the discussion in \citet{Abdollahi}), the uncertainties of these two data points would be much larger and they would agree with the DAMPE data and our result.

For dark matter annihilating via the $\mu^+\mu^-$ and $\tau^+\tau^-$ channels, we can see that these two channels are difficult to produce the required flux (see Fig.~4). For the other quark channels (e.g. $b\bar{b}$) and $W^+W^-$ channel, the maximum flux position is near 1-10 GeV, which are not able to match the 1.4 TeV peak. Therefore, only the $e^+e^-$ channel can produce the required peak.

\section{Discussion}
We propose a new simple model to account for the 1.4 TeV peak in the DAMPE data. The enhanced dark matter annihilation via the $e^+e^-$ channel surrounding our closest black hole (A0620-00) can give the right amount of electron and positron flux to account for the 1.4 TeV peak. The predicted dark matter mass is 2.75 TeV (with MED model parameters) and we have assumed the thermal relic annihilation cross section. Generally speaking, $m \sim 1.5-3$ TeV can give very good agreement with the DAMPE data, with the uncertainty of the diffusion parameters. Only one free parameter, $t_{\rm BH}$, is involved in the calculations. It is difficult to determine the actual age of the black hole. Here, we have assumed $t_{\rm BH}=5$ Gyr, which is a reasonable estimation and matches the age of the disk in our Milky Way. If we take $t_{\rm BH}=3-8$ Gyr, the resulting flux is still within the uncertainty of the data point. Therefore, our simple model can give a satisfactorily explanation to the observed peak in the DAMPE data. 

Some previous observations and analyses have provided stringent constraints of $m$ for annihilating dark matter. For instance, the positron and electron detections from AMS-02 have excluded $m \le 90$ GeV for thermal relic dark matter annihilating via $e^+e^-$ channel \citep{Bergstrom}. Besides, the upper limit of annihilation cross section for $m=1.5-3$ TeV annihilating via $\mu^+\mu^-$ channel is $\sigma v \sim 10^{-25}$ cm$^3$ s$^{-1}$ based on the H.E.S.S. gamma-ray observations \citep{Abdallah}. This upper limit also approximately applies for the $e^+e^-$ channel as the total amount of gamma-ray emitted from the $e^+e^-$ channel is close to that from the $\mu^+\mu^-$ channel. Therefore, our result on the mass range $m=1.5-3$ TeV with the thermal relic annihilation cross section $\sigma v=2.2 \times 10^{-26}$ cm$^3$ s$^{-1}$ can evade the current stringent limits on $m$ and $\sigma v$.

In fact, there is one known black hole (in Cygnus X-1) with mass $14.81 \pm 0.98M_{\odot}$ located at $1.86^{+0.12}_{-0.11}$ kpc from us (the second nearest known black hole) \citep{Orosz}. However, since the diffusion distance is longer, the required dark matter mass would be much larger. In order to produce the 1.4 TeV peak, the required dark matter mass is 21 TeV, which would largely suppress the electron/positron flux ($\phi_{\rm sp} \propto m_{\rm TeV}^{-4/3}$). Therefore, if there exists some unknown black holes which are located at distance smaller than 1 kpc, they can significantly contribute to the resultant electron/positron flux. However, these potential flux contributions depend on the black hole mass and the actual distance, which are difficult to model and predict. Based on the fact that we have not yet discovered any black hole within 1 kpc from us, we have omitted these potential flux contributions.

Previous studies mainly focus on assuming that the 1.4 TeV peak originates from dark matter decay or annihilation from a nearby dark matter subhalo or ultra-compact micro halos \citep{Pan,Jin,Athron,Huang,Yang,Ding}. However, there is no strong evidence showing the existence of dark matter subhalos or compact halos in our galaxy and the results obtained from these studies are very uncertain. In our model, we rely on the fact of the existence of a nearby black hole and the standard physics of the dark matter spike formation. We use the thermal relic annihilation cross section in standard cosmology and follow the standard diffusion scenario in high-energy astrophysics to calculate the predicted dark matter mass. Using a reasonable value of $t_{\rm BH}$, the predicted electron and positron flux gives an excellent agreement with the DAMPE data. Therefore, compared with the previous studies, our model is an outstanding one (with the fewest assumptions) to explain the 1.4 TeV peak.

There are three important implications based on this result. First, we can obtain the mass of dark matter, which is about 1.5-3 TeV. This mass range can be severely tested in future collider experiments. Second, dark matter would annihilate via $e^+e^-$ channel with a very large branch ratio. Similar signals can also be detected in dark matter dominated galaxies (e.g. Milky Way dwarf spheroidal satellite galaxies), which can be verified by future high-energy observations (e.g. radio observations). Third, the signal of dark matter annihilation can be greatly enhanced by black holes. This has been discussed extensively about the emission near a supermassive black hole \citep{Fields,Lacroix2}. More studies in this direction can verify our model and disclose the mystery of dark matter.

\begin{table}
\caption{Benchmark propagation model parameters in our galaxy \citep{Delahaye}.}
 \label{table1}
 \begin{tabular}{@{}lcc}
  \hline
  Model &  $K_0$ (kpc$^2$/Myr) & $\delta$ \\
  \hline
  MIN & 0.00595 & 0.55 \\
  MED & 0.0112 & 0.70 \\
  MAX & 0.0765 & 0.46 \\
  \hline
 \end{tabular}
\end{table}

\begin{figure}
\vskip 10mm
 \includegraphics[width=80mm]{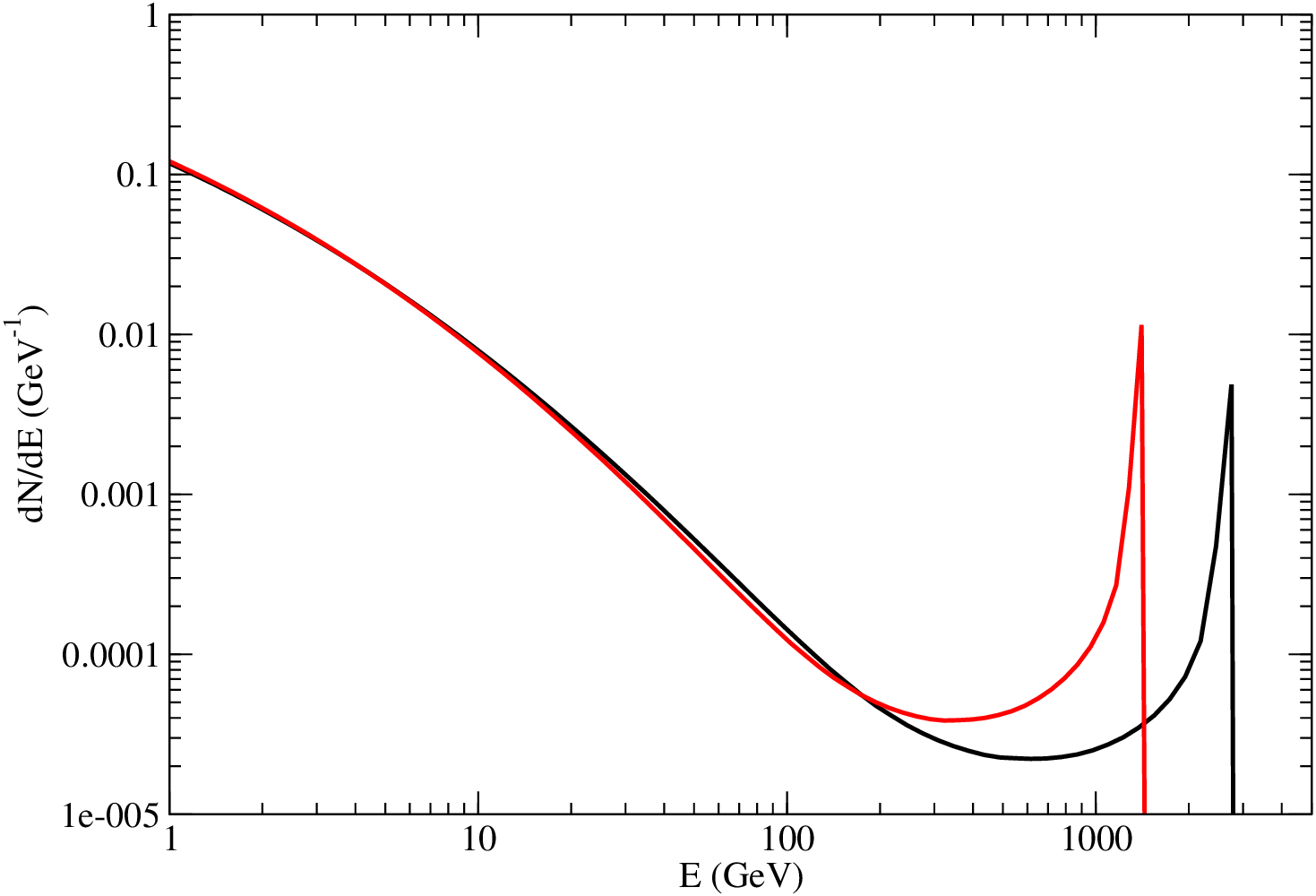}
 \caption{The electron energy spectra before (black line) and after (red line) propagation, assuming the MED model parameters and $m=2.75$ TeV.}
\vskip 10mm
\end{figure}

\begin{figure}
\vskip 10mm
 \includegraphics[width=80mm]{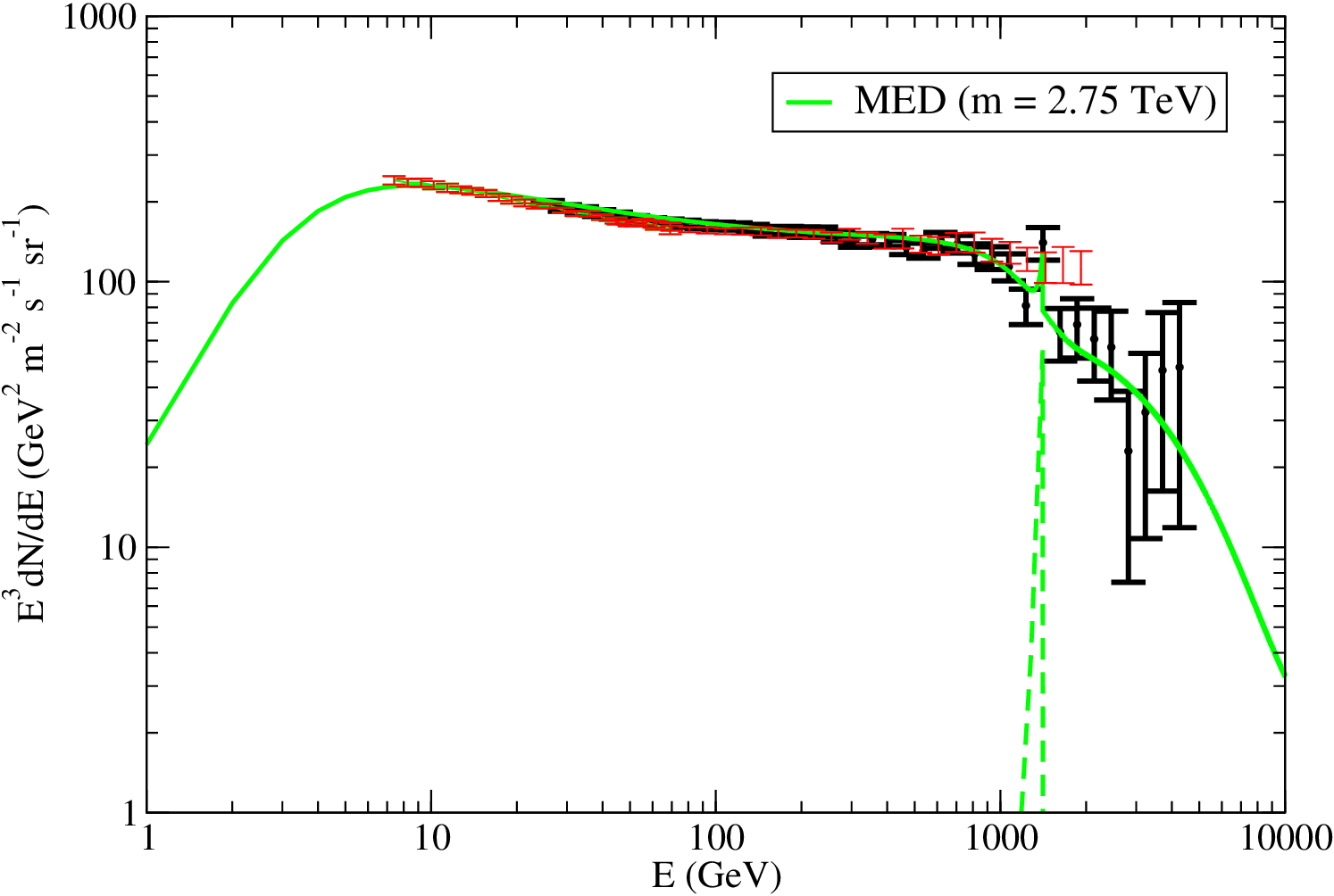}
 \caption{The flux spectrum $\phi$ of our model assuming MED model parameters. The green dashed line represents the contribution of the spike emission $\phi_{\rm sp}$ (via $e^+e^-$ channel). The black dots with error bars are the observed DAMPE spectrum taken from \citet{Ambrosi}. The red data points with error bars are the Fermi-LAT spectrum \citep{Abdollahi}.}
\vskip 10mm
\end{figure}

\begin{figure}
\vskip 10mm
 \includegraphics[width=80mm]{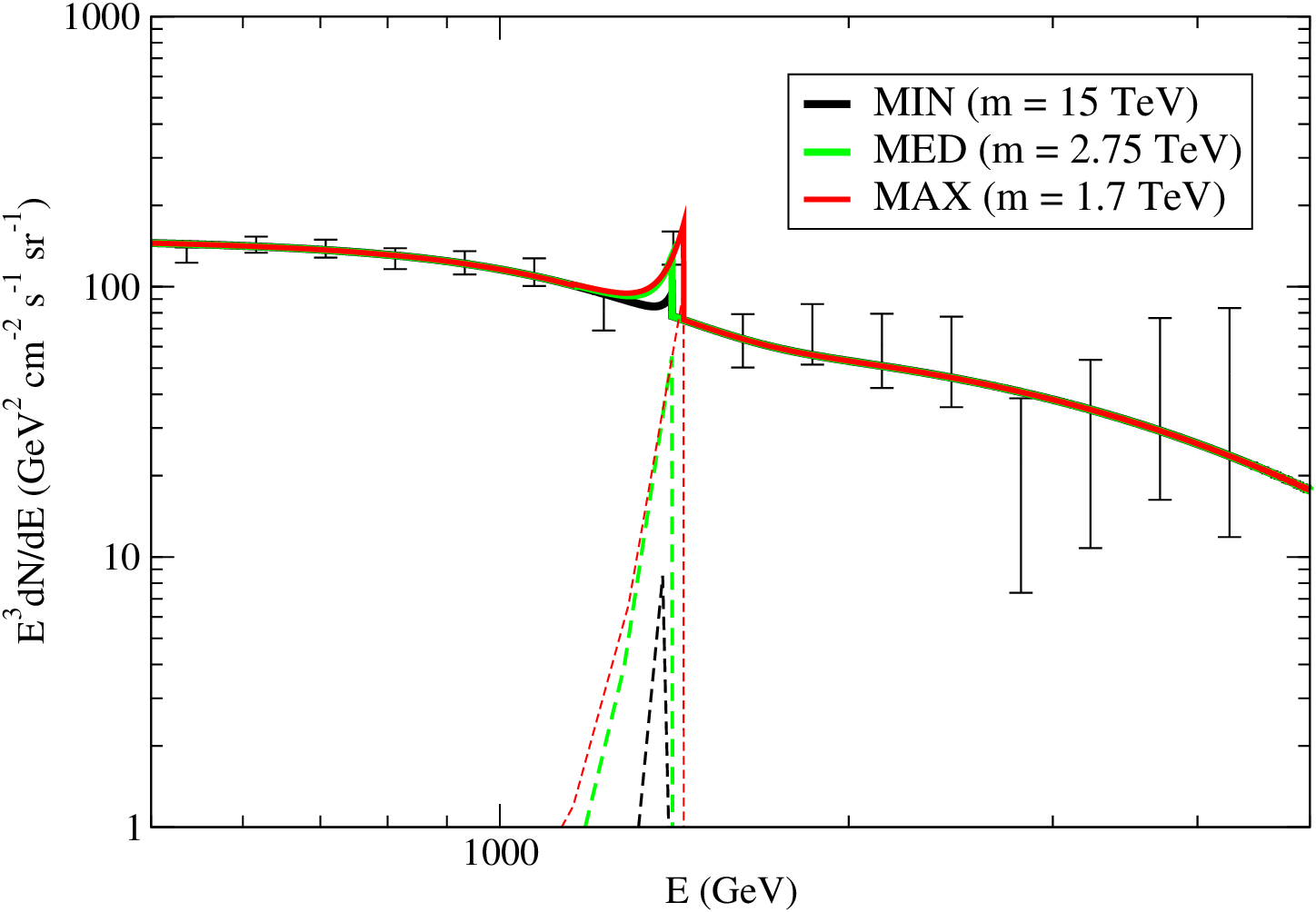}
 \caption{The flux spectra $\phi$ of our model assuming MIN, MED and MAX model parameters. The dashed lines represent the contributions of the spike emission $\phi_{\rm sp}$ for corresponding model parameters. The data are taken from \citet{Ambrosi}.}
\vskip 10mm
\end{figure}

\begin{figure}
\vskip 10mm
 \includegraphics[width=80mm]{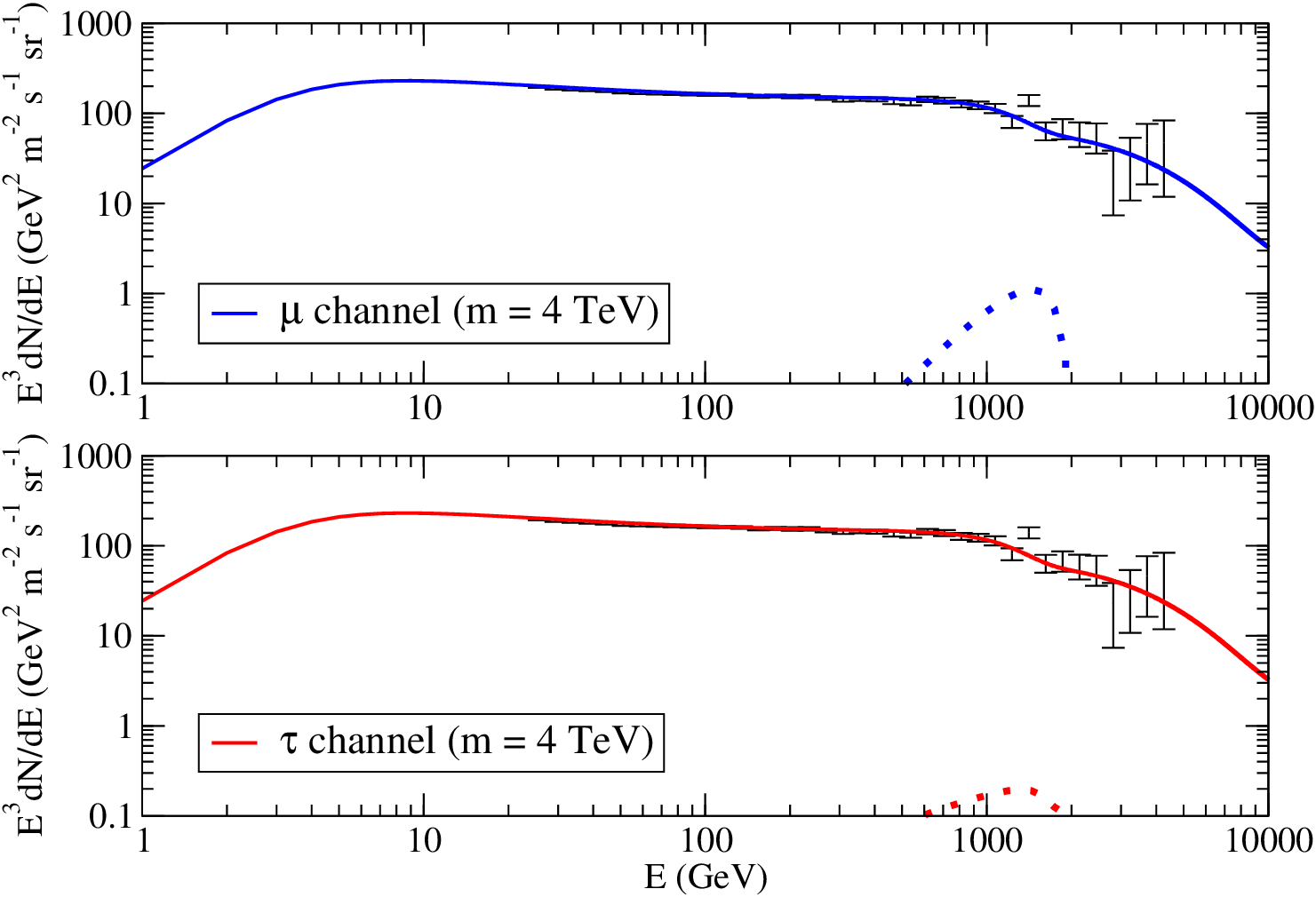}
 \caption{The flux spectra $\phi$ of our model assuming MED model parameters. The dotted lines represent the contributions of the spike emission $\phi_{\rm sp}$ (blue: $\mu^+\mu^-$ channel; red: $\tau^+\tau^-$ channel). The data are taken from \citet{Ambrosi}.}
\vskip 10mm
\end{figure}

\section{acknowledgements}
This work was supported by a grant from the Research Grants Council of the Hong Kong Special Administrative Region, China (Project No. EdUHK 28300518).

\label{lastpage}

\end{document}